\documentclass[a4paper,12pt]{article}
\usepackage{amssymb,amsthm,latexsym}
\newcommand{\Med}[1]{\left\langle #1 \right\rangle}
\newcommand{\med}[1]{\langle #1 \rangle}

\newtheorem{theorem}{Theorem}
\newtheorem{lemma}{Lemma}

\title{The Thermodynamic Limit in Mean Field Spin Glass Models}

\author{
Francesco Guerra\footnote{\
e-mail: {\tt francesco.guerra@roma1.infn.it}} \\
{\small {\itshape Dipartimento di Fisica, Universit\`a di Roma `La Sapienza'}}
\\
{\small {\itshape and INFN, Sezione di Roma, Piazzale A. Moro 2, 00185 Roma, 
Italy}}\\
Fabio Lucio Toninelli\footnote{\ 
e-mail: {\tt f.toninelli@sns.it}} \\
{\small {\itshape Scuola Normale Superiore, Piazza dei Cavalieri 7, 56126 Pisa,
Italy}}\\
{\small {\itshape and Istituto Nazionale di Fisica Nucleare, Sezione di 
Pisa}}
} 

\date{\today}

\begin{document}

\maketitle

\begin{abstract}
We present a simple strategy in order to show the existence and uniqueness of 
the infinite volume limit of thermodynamic quantities, 
for a large class of mean field disordered 
models, as for example the Sherrington-Kirkpatrick model, and the Derrida 
p-spin model. The main argument is based on a smooth interpolation between a 
large system, made of $N$ spin sites, and two similar but 
independent subsystems, made 
of $N_1$ and $N_2$ sites, respectively, with $N_1+N_2=N$. The quenched 
average of the free energy turns out to be subadditive with respect to the 
size of the system. This gives immediately 
convergence of the free energy per site, in the infinite volume limit. 
Moreover, a simple argument, based on concentration of measure, gives the 
almost sure convergence, with respect to the external noise. Similar 
results hold also for the ground state energy per site. 
\end{abstract}

\newpage

\section{Introduction}

The main objective of this paper is to propose a general strategy in order to 
control the infinite volume limit of thermodynamic quantities for a class 
of mean 
field spin glass models. 
For the sake of definiteness, we consider firstly in full detail
the Sherrington-Kirkpatrick 
(SK) 
model, \cite{sk}, \cite{MPV}.
Then, we 
show how to generalize our method to similar related mean field 
disordered models, as for example the Derrida p-spin model, 
\cite{derrida}, \cite{gross}.

It is very well known that the rigorous control of the infinite volume limit for 
these mean field models is very difficult, due to the effects of very 
large fluctuations produced by the external noise. In particular, it is 
very difficult to produce very effective trial states, to be exploited in 
variational principles. It is only for the high temperature, or high 
external field, regime that a satisfactory control can be reached, as 
shown for example in 
\cite{shch}, \cite{talaHT}, \cite{quadr}.

We will introduce a very simple strategy for the control of the infinite 
volume limit. The main idea is to split a large system, made of $N$ spin 
sites, into two subsystems, made of $N_1$ and $N_2$ sites, respectively, 
where
each subsystem is subject to some external noise, similar but {\it independent}
from the noise acting on the large system. By a smooth interpolation 
between the system and the subsystems, we will show subadditivity 
of the quenched average of the free energy, with respect to the size of the 
system, and, therefore, obtain a complete control of the infinite volume limit.

Moreover, the well known selfaveraging of the free energy density, as shown 
originally by Pastur and Shcherbina in \cite{pastur}, extended to the estimates 
given by the concentration of measure, as explained in \cite{T}  and 
\cite{L}, does allow an even more detailed control of the limit. In effect, 
it will turn out that the free energy per site, without quenched average, 
converges almost surely, with respect to the external noise. These results 
extend to other thermodynamic quantities, in particular to the ground state 
energy per site, as it will be shown in the paper.

The organization of the paper is as follows. In Section 2 we  recall 
the general structure of the Sherrington-Kirkpatrick 
mean field spin glass model, in order to define the main quantities, and
fix the notations. Next Section 3 contains the main results of the paper, 
related to the control of the infinite volume limit. In Section 4 we show 
how to extend our results to other mean field spin glass models, in 
particular to the Derrida p-spin model and to models with non-Gaussian 
couplings. Section 5 contains conclusions and outlook for 
future developments and extensions.

\section{The structure of the Sherrington-Kirkpatrick model}

Let us recall some basic definitions.
 
Ising spin variables
$\sigma_{i}=\pm 1$,  attached to each site $i=1,2,\dots,N$, 
define the generic configuration of the mean field spin glass model.
The external quenched disorder is
given by the $N(N-1)/2$ independent and identical distributed random
variables $J_{ij}$, defined for each couple of sites. For the sake of simplicity,
we assume each $J_{ij}$ to be a centered
unit Gaussian with averages
$$E(J_{ij})=0,\quad E(J_{ij}^2)=1.$$
The Hamiltonian of the model, in some external field of strength $h$,  
is given by
\begin{equation}\label{H}
H_N(\sigma,h,J)=-{1\over\sqrt{N}}\sum_{1\le i<j\le N}J_{ij}\sigma_i\sigma_j
-h\sum_{i=1}^N\sigma_i.
\end{equation}
The first term in (\ref{H}) is a long range random two body interaction,
while the second represents the interaction of the spins with a fixed 
external magnetic field $h$.

For a given inverse temperature $\beta$, we introduce the disorder dependent
partition function $Z_{N}(\beta,h,J)$, 
the quenched average of the free energy per site
$f_{N}(\beta,h)$, the Boltzmann state 
$\omega_J$, and the auxiliary function $\alpha_N(\beta,h)$,  
according to the definitions
\begin{eqnarray}\label{Z}
&&Z_N(\beta,h,J)=\sum_{\{\sigma\}}\exp(-\beta H_N(\sigma,h,J)),\\
\label{f}
&&-\beta f_N(\beta,h)=N^{-1} E\log Z_N(\beta,h,J)=\alpha_N(\beta,h),\\
\label{state}
&&\omega_{J}(A)=Z_N(\beta,h,J)^{-1}\sum_{\{\sigma\}}A\exp(-\beta
H_N(\sigma,h,J)), 
\end{eqnarray}
where $A$ is a generic function of the $\sigma$'s. In the
notation $\omega_J$, we have stressed the dependence of the Boltzmann
state on the external noise $J$, but, of course, there is also
a dependence on $\beta$, $h$ and $N$.

Let us now introduce the important concept of replicas. 
Consider a generic number $s$ of independent copies
of the system, characterized by the Boltzmann
variables $\sigma^{(1)}_i$, $\sigma^{(2)}_i$, $\dots$,
distributed according to the product state
$$\Omega_J=\omega^{(1)}_J \omega^{(2)}_J \dots\omega^{(s)}_J,$$
where all $\omega^{(\alpha)}_J$ act on each one
$\sigma^{(\alpha)}_i$'s, and are subject to the {\sl
same} sample $J$ of the external noise. Clearly, the {\sl Boltzmannfaktor} for the replicated system is given by
\begin{equation}
\exp\left(-\beta
(H_N(\sigma^{(1)},h,J)+H_N(\sigma^{(2)},h,J)+\dots+H_N(\sigma^{(s)},h,J))
\right).
\end{equation}
The overlaps between any two replicas $a,b$ are
defined according to
$$q_{ab}(\sigma^{(a)},\sigma^{(b)})={1\over
N}\sum_{i}\sigma^{(a)}_i\sigma^{(b)}_i,$$
and they satisfy the obvious bounds
$$-1\le q_{ab}\le 1.$$

For a generic smooth function $F$ of the overlaps, we
define the $\langle\rangle$ averages
$$\langle F(q_{12},q_{13},\dots)\rangle=E\Omega_J\bigl(F(q_{12},q_{13},\dots)\bigr),$$
where the Boltzmann averages $\Omega_J$ acts on the replicated $\sigma$
variables, and $E$ is the average with respect to the external noise $J$.

\section{Control of the infinite volume limit}

Let us explain the main idea behind our method.
We divide the $N$ sites into two blocks $N_1,N_2$ with $N_1+N_2=N$,
and define 
\begin{eqnarray}
\label{Z_N^t}
Z_N(t)&=&\sum_{\{\sigma\}}\exp\left(\beta\sqrt{\frac tN}\sum_{1\le i<j\le N}
J_{ij}\sigma_i\sigma_j+\beta\sqrt{\frac {1-t}{N_1}}\sum_{1\le i<j\le N_1}
J'_{ij}\sigma_i\sigma_j\right.\\\nonumber
&&\left.+\beta\sqrt{\frac {1-t}{N_2}}\sum_{N_1 <i<j\le N}
J''_{ij}\sigma_i\sigma_j\right)\exp \beta h\sum_{i=1}^N\sigma_i,
\end{eqnarray}
with $0\le t\le 1$.

The external noise is represented by the {\it independent} families of 
unit Gaussian random
variables $J$, $J'$ and $J''$. Notice that the two subsystems are subject 
to a different 
external noise, with respect to the original system. But, of course, the 
probability distributions are the same.
The parameter $t$ allows to interpolate between the original $N$ spin system
at $t=1$ and a system composed of two non interacting parts at $t=0$, so 
that
\begin{eqnarray} 
&&Z_N(1)=Z_N(\beta,h,J)\\
&&Z_N(0)=Z_{N_1}(\beta,h,J')Z_{N_2}(\beta,h,J'').
\end{eqnarray}
As a consequence, by taking into account the definition in (\ref{f}), we 
have
\begin{eqnarray}
\label{boundary}
&& E\ln Z_N(1)=N\alpha_N(\beta,h)\\
&& E\ln Z_N(0)={N_1}\alpha_{N_1}(\beta,h)
+{N_2}\alpha_{N_2}(\beta,h).
\end{eqnarray}
By taking the derivative of $N^{-1} E\ln Z_N(t)$ with respect to the 
parameter $t$, we obtain
\begin{eqnarray}
\label{derivata}
&&\frac d{dt}\frac1N E\,\ln Z_N(t)=\frac{\beta}{2N}E\left(
\frac1{\sqrt {tN}}\sum_{1\le i<j\le N}J_{ij}\omega_t(\sigma_i\sigma_j)
\right.\\\nonumber
&&\left.-\frac1{\sqrt {(1-t)N_1}}
\sum_{1\le i<j\le N_1}J'_{ij}\omega_t(\sigma_i\sigma_j)
-\frac1{\sqrt {(1-t)N_2}}
\sum_{N_1< i<j\le N}J''_{ij}\omega_t(\sigma_i\sigma_j)
\right)
\end{eqnarray}
where $\omega_t(.)$ denotes the Gibbs state corresponding to the
partition function (\ref{Z_N^t}).
A standard integration by parts on the Gaussian noise, as done for example in 
\cite{MPV}, \cite{guerra2}, gives
\begin{eqnarray}
\nonumber
\frac d{dt}\frac1N E\,\ln Z_N(t)&=&\frac{\beta^2}{4N^2}\sum_{i,j=1}^N
E(1-\omega_t^2(\sigma_i\sigma_j))-\frac{\beta^2}{4N N_1}\sum_{i,j=1}^{N_1}
E(1-\omega_t^2(\sigma_i\sigma_j))
\\
\label{dt}
&&-\frac{\beta^2}{4N N_2}\sum_{i,j=N_1+1}^N
E(1-\omega_t^2(\sigma_i\sigma_j))\\
&=&
-\frac{\beta^2}4\Med{q_{12}^2-\frac{N_1}N(q_{12}^{(1)})^2-
\frac{N_2}N(q_{12}^{(2)})^2},
\end{eqnarray}
where we have defined  
\begin{eqnarray}
&&N_1\, q_{12}^{(1)}=\sum_{i=1}^{N_1}\sigma^1_i\sigma^2_i\\
&&N_2\, q_{12}^{(2)}=\sum_{i=N_1+1}^{N}\sigma^1_i\sigma^2_i.
\end{eqnarray}
Since $q_{12}$ is a convex linear combination of $q_{12}^{(1)}$ 
and $q_{12}^{(2)}$ in the form
$$
q_{12}=\frac{N_1}N\, q_{12}^{(1)}+ \frac{N_2}N\, q_{12}^{(2)},
$$
due to convexity of the function $f: x\rightarrow x^2$, we have the 
inequality
$$
\Med{q_{12}^2-\frac{N_1}N(q_{12}^{(1)})^2-\frac{N_2}N(q_{12}^{(2)})^2}\leq 0.
$$

Therefore, we can state our first preliminary result.
\begin{lemma}
\label{increase}
The quenched average of the logarithm of the interpolating partition 
function, defined by (\ref{Z_N^t}), is increasing in $t$, {\it i.e.}
\begin{equation}
\frac d{dt}\frac1N E\,\ln Z_N(t)\geq 0.
\end{equation}
\end{lemma}
By integrating in $t$ and recalling the boundary conditions
(\ref{boundary}), we get the first main result.
\begin{theorem}
\label{super}
The following superadditivity property holds 
\begin{equation}
\label{superadd}
N\, \alpha_N(\beta,h)\geq N_1\, \alpha_{N_1}(\beta,h)+
N_2\, \alpha_{N_2}(\beta,h).
\end{equation}
Of course, due the minus sign in (\ref{f}), we have subadditivity for the 
quenched average of the free energy.
\end{theorem}

The subadditivity property gives an immediate control on the infinite 
volume limit, as explained for example in \cite{ruelle}. In fact, we have
\begin{theorem}
\label{limit}
The infinite volume limit for $\alpha_N(\beta,h)$ does exists and equals its
$\sup$ 
\begin{equation}
\label{risultato1}
\lim_{N\to\infty}\alpha_N(\beta,h)=\sup_N \alpha_N(\beta,h)\equiv 
\alpha(\beta,h).
\end{equation}
\end{theorem}

For finite $N$ and a given realization $J$ of the disorder, define the 
ground state energy density $-e_N(J,h)$ as
\begin{equation}
\label{eN}
-e_N(J,h)=\frac1N\, \inf_{\sigma}H_N(\sigma,h,J).
\end{equation}

Now we show, from simple thermodynamic properties, that Eq. (\ref{risultato1})
implies the existence of the thermodynamic limit for $E\,(e_N(h,J))$.
First of all, notice that the bounds
\begin{equation}
e^{\beta N e_N(J,h)}\leq \sum_{\{\sigma\}}e^{-\beta H_N(\sigma,h,J)}\leq
2^Ne^{\beta N e_N(J,h)}
\end{equation}
hold for any $J,N,\beta,h$, so that
\begin{equation}
\label{bounds}
0\le \frac{\ln Z_N(\beta,h,J)}{\beta N}-e_N(J,h)\le \frac{\ln 2}\beta.
\end{equation}
The bounds (\ref{bounds}), together with the obvious  
$$
\partial_\beta \frac{\ln Z_N(\beta,h,J)}{\beta }\le 0,
$$
imply that
\begin{equation}
\lim_{\beta\to\infty}\frac{\ln Z_N(\beta,h,J)}{\beta N}\downarrow e_N(J,h).
\end{equation}
Of course, by taking the expectation value in (\ref{bounds}) and defining
$$
e_N(h)=E\,e_N(J,h),
$$
one also finds
\begin{equation}
\label{ae}
\lim_{\beta\to\infty}\frac{\alpha_N(\beta,h)}{\beta N}\downarrow e_N(h).
\end{equation}
Therefore, by taking into account the superadditivity  
(\ref{superadd}), the inequalities (\ref{bounds}), and 
the existence of the limit $\alpha(\beta,h)$ for
$\alpha_N(\beta,h)$, we have from (\ref{ae}) the proof of the following 
\begin{theorem}
\label{lime1}
For the quenched average of the ground state energy we have the
subadditivity property 
\begin{equation}
N\, e_N(h)\geq N_1\, e_{N_1}(h)+N_2\, e_{N_2}(h).
\end{equation}
and the existence of the infinite volume limit
\begin{equation}
\lim_{N\to\infty} e_N(h)=\sup_N e_N(h)\equiv e_0(h).
\end{equation}
Finally,  we can  write the limit $e_0(h)$ in terms of $\alpha(\beta,h)$ 
as
\begin{equation}
\label{e0}
\lim_{\beta\to\infty}\frac{\alpha(\beta,h)}{\beta}\downarrow e_0(h).
\end{equation}
\end{theorem}

%
%

After proving the existence of the thermodynamic limit 
for the quenched averages, we can easily extend our results to prove 
that convergence holds for almost every disorder realization $J$.
In fact, we can state
\begin{theorem}
\label{as}
The infinite volume limits
\begin{equation}
\label{lima}
\lim_{N\to\infty}\frac 1N \ln Z_N(\beta,h,J)=\alpha(\beta,h),
\end{equation}
\begin{equation}
\label{lime}
\lim_{N\to\infty}e_N(J,h)=e_0(h),
\end{equation}
do exist $J$-almost surely.
\end{theorem}

For the proof, we notice that the fluctuations of the free energy 
per site 
vanish exponentially fast as $N$ grows, a result strengthening the 
pioneering quadratic selfaveraging proven in \cite{pastur}. 
Indeed, the following result 
holds \cite{T}
\begin{equation}
\label{fluttuazioni}
P\left(\left|\frac 1{\beta N}\ln Z_N(\beta,h,J)-
\frac 1{\beta N}E\,\ln Z_N(\beta,h,J)\right|\ge u\right)\le e^{-N{u^2}/2}.
\end{equation}
Since the r.h.s. of (\ref{fluttuazioni}) is summable in $N$ for every
fixed $u$, Borel-Cantelli lemma \cite{shiri}, and the convergence given 
by (\ref{risultato1}) imply (\ref{lima}).
The same argument can be exploited for the ground state energy. In fact,
by taking the $\beta\to\infty$ limit in (\ref{fluttuazioni}), we get  
\begin{equation}
\nonumber
P\left(\left|e_N(J,h)-e_N(h)\right|\ge u\right)\le e^{-N{u^2}/2}.
\end{equation}
Again, Borel-Cantelli lemma implies (\ref{lime}), and the Theorem is 
proven.

Notice that all the results of this Section hold also in the case where
on each spin $\sigma_i$ acts a random magnetic field $h_i$, where
the $h_i$'s are i.i.d. random variables.

%
%

\section{Existence of the thermodynamic limit for other mean
field spin glass models}

In this Section, we show how the above results on the (almost sure)
existence of the thermodynamic limit for the free energy and for the ground
state energy can be extended to other mean field spin glass models. 
In the first place, we can immediately extend the approach to $p$-spin 
models, for even $p$.
On the other hand, we can allow the quenched disorder variables 
to be non-Gaussian, 
provided that suitable bounds are imposed on their moments.

\subsection{$p$-spin models}

%
%

The $p$-spin model is defined by the Hamiltonian
\begin{eqnarray}
H^{(p)}_N(\sigma,h,J)=
-\sqrt\frac{p!}{2 N^{p-1}}\sum_{(i_1,\ldots,i_p)} J_{i_1\ldots
i_p} \sigma_{i_1}\ldots\sigma_{i_p}-h\sum_i \sigma_i,
\end{eqnarray}
where $p$ is an integer and $J_{i_1\ldots i_p}$ are i.i.d. unit Gaussian 
random variables. The summation in the first term is performed on all the
different $p$-ples of indices $i_1,\ldots,i_p$.
Note that for $p=2$ this is just the SK model.
The $p$-spin model has been proposed by Derrida in \cite{derrida}, and
extensively studied thereafter, see for example 
\cite{gross}, \cite{MPV}, \cite{talapspin1}, 
\cite{talapspin2}.

For the sake of simplicity, we consider the case of even $p$.
As we did for the SK model, we define the auxiliary partition function 
$Z_N(t)$, in 
analogy with (\ref{Z_N^t}). By taking the $t$ derivative, we
find after integration by parts
\begin{eqnarray}
\label{pspin}
\frac d{dt}\frac1N E\,\ln Z_N(t)\!\!\!&=&\!\!\!
-\frac{\beta^2}4\Med{q_{12}^p-\frac{N_1}N(q_{12}^{(1)})^p-
\frac{N_2}N(q_{12}^{(2)})^p}+O(1/N)\\\nonumber
&&\geq O(1/N),
\end{eqnarray}
for $p$ even, by the same convexity argument as before. 
It is easy to realize the reason for the appearance of the terms $O(1/N)$.
In fact, for $p=2$, we can write
\begin{equation}
\frac2{N^2}
\sum_{(i,j)}E\,(1-\omega_t^2(\sigma_{i}\sigma_{j}))=
\frac{1}{N^2}
\sum_{i,j=1}^NE\,(1-\omega_t^2(\sigma_{i}\sigma_{j}))=(1-\med{q_{12}^2}),
\end{equation}
as already exploited in (\ref{dt}). On the other hand, 
for $p>2$ one has
\begin{eqnarray}
\nonumber
\frac{p!}{N^p}
\sum_{(i_1,\ldots,i_p)}E\,(1-\omega_t^2(\sigma_{i_1}\ldots\sigma_{i_p}))&=&
\frac{1}{N^p}
\sum_{i_1,\ldots,i_p=1}^NE\,(1-\omega_t^2(\sigma_{i_1}\ldots\sigma_{i_p}))+
O(1/N)\\
&=&(1-\med{q_{12}^p})+O(1/N).
\end{eqnarray}
From (\ref{pspin}) one finds, as in the previous Section, the existence of 
the infinite volume limits
\begin{eqnarray}
&&\lim_{N\to\infty}\alpha^{(p)}_N(\beta,h)\equiv \alpha^{(p)}
(\beta,h)\\
&&\lim_{N\to\infty}e_N^{(p)}(h)\equiv e^{(p)}_0(h)
=\lim_{\beta\to\infty}\frac{\alpha^{(p)}(\beta,h)}\beta.
\end{eqnarray}
Moreover, the estimate (\ref{fluttuazioni}) holds also in this case,
since it is based only on the fact that $1/N\beta\ln Z_N(\beta,h,J)$, as 
a function of the variables $J$, has a Lipshitz constant of order $C/\sqrt N$,
where $C$ is a constant independent of $\beta$ \cite{T}.
Therefore, also in this case
we have almost sure convergence for the free energy and for the
ground state energy.

\subsection{Non-Gaussian couplings}

\label{non-gauss}

The method developed in the previous Sections allows to prove the 
existence of the thermodynamic limit 
for $\alpha_N(\beta,h)$ and for $e_N(h)$ also
for SK (or $p$-spin) models with non-Gaussian couplings, 
provided that the variables $J_{ij}$ (respectively, $J_{i_1,\ldots,i_p}$)
are i.i.d. symmetric random variables with finite fourth moment, i.e., 
$P(J_{ij})=P(-J_{ij})$
and $E J_{ij}^4<\infty$. A similar condition has been also exploited for 
the study of the model, at zero external field and high temperature, 
see for example \cite{alr} and \cite{fro1}.
Consider for instance the SK case.
The integration by parts on the disorder variables in (\ref{derivata}) can
be performed by use of the formula
\begin{equation}
\label{perparti}
E\,\eta F(\eta)=E\,\eta^2F'(\eta)-\frac14E\,|\eta|\int_{-|\eta|}^{|\eta|}
(\eta^2-t^2)F'''(t)\,dt,
\end{equation}
which holds for any symmetric random variable $\eta$ and for sufficiently
regular functions $F$, as a simple direct calculation shows. 
A similar expression has been exploited by Talagrand in \cite{talacergy}, 
for dicotomic variables.
By applying this formula to the various terms in (\ref{derivata}), one 
finds that
\begin{equation}
\frac d{dt}\frac1N E\,\ln Z_N(t)=-\frac{\beta^2J^2}4
\Med{q_{12}^2-\frac{N_1}N(q_{12}^{(1)})^2-\frac{N_2}N(q_{12}^{(2)})^2}
+O(N^{-1/2}),
\end{equation}
where 
$$
J^2=E\,J_{ij}^2.
$$
The error terms arise from the estimates 
$$
\partial^3_{J_{ij}}\omega_t(\sigma_i\sigma_j)=O(N^{-3/2})
$$
and 
$$
E\, J_{ij}^2\omega^2_t(\sigma_i\sigma_j)=J^2 E\,\omega^2_t(\sigma_i\sigma_j)
+O(N^{-1/2}).
$$
The existence of the thermodynamic limit for the quenched averages of the
free energy and the 
ground state energy then follows.

In order to prove $J$-almost sure convergence, for the sake 
of simplicity, let us consider the case where  the random variables 
$J_{ij}$ are bounded. Then, the estimate (\ref{fluttuazioni}) still holds, 
with the r.h.s. modified into
$\exp(-N\,K\,u^2)$, where $K$ does not depend on $\beta$.
This can be proved, for instance, by using Theorem 6.6 of \cite{newlook}.
Almost sure convergence for the free energy and for the ground state energy
then follows immediately from  Borel-Cantelli lemma.
In particular, this includes
the important case where $J_{ij}=\pm1$ with equal probability. The 
extension to more general cases, where only the condition $E J_{ij}^4<\infty$
is required, is possible, and will be reported in future work.

\section{Conclusions and outlook}

We have seen that the rigorous control of the infinite volume limit 
for mean field 
spin glass models can be obtained through a simple strategy, by a smooth 
interpolation between a large system and its splitting into subsystems, 
provided the external noises are taken independent. 

The extension of our methods to the important cases of diluted models, 
and other models of the neural 
network type, will be taken into account in future work.

\vspace{.5cm}
{\bf Acknowledgments}

We gratefully acknowledge useful conversations with
Pierluigi Contucci, Roberto D'Autilia, Flora Koukiou, Enzo Marinari, 
Giorgio Parisi, Masha Scherbina, and Michel Talagrand.

This work was supported in part by MIUR 
(Italian Minister of Instruction, University and Research), 
and by INFN (Italian National Institute for Nuclear Physics).

\end{document}